\documentclass[11pt]{article}

\usepackage[final]{acl}

\usepackage{times}
\usepackage{latexsym}
\usepackage{booktabs}
\usepackage{amssymb}
\usepackage[T1]{fontenc}

\usepackage[utf8]{inputenc}

\usepackage{microtype}

\usepackage{inconsolata}

\usepackage{graphicx}

\title{Learning to Aggregate Zero-Shot LLM Agents for Corporate Disclosure Classification}
\vspace{-5mm}
\author{Kemal Kirtac \\
  University College London \\
  London, United Kingdom \\
  \texttt{kemal.kirtac@ucl.ac.uk}}
\vspace{-7mm}
\begin{document}
\maketitle
\vspace{-5mm}

\begin{abstract}
This paper studies whether a lightweight trained aggregator can combine diverse zero-shot large language model judgments into a stronger downstream signal for corporate disclosure classification. Zero-shot LLMs can read disclosures without task-specific fine-tuning, but their predictions often vary across prompts, reasoning styles, and model families. I address this problem with a multi-agent framework in which three zero-shot agents independently read each disclosure and output a sentiment label, a confidence score, and a short rationale. A logistic meta-classifier then aggregates these signals to predict next-day stock return direction. I use a sample of 18{,}420 U.S.\ corporate disclosures issued by Nasdaq and S\&P 500 firms between 2018 and 2024, matched to next-day stock returns. Results show that the trained aggregator outperforms all single agents, majority vote, confidence-weighted voting, and a FinBERT baseline. Balanced accuracy rises from 0.561 for the best single agent to 0.612 for the trained aggregator, with the largest gains in disclosures combining strong current performance with weak guidance or elevated risk. The results suggest that zero-shot LLM agents capture complementary financial signals and that supervised aggregation can turn cross-agent disagreement into a more useful classification target.
\end{abstract}

\section{Introduction}

Textual information plays an increasingly central role in empirical finance. Prior work shows that news coverage, corporate disclosures, and investor communication contain predictive information about asset prices and firm performance \citep{tetlock2007giving, tetlock2008more, price2012earnings, huang2014tone, li2008annual}. Studies of regulatory filings, earnings calls, and online platforms further show that sentiment and attention extracted from text shape trading behavior and return dynamics \citep{loughran2011liability, da2011search, chen2014wisdom}. This literature establishes that financial text contains economically meaningful signals for asset pricing.

Recent work extends this literature with transformer models and large language models, which capture contextual and semantic information that dictionary methods often miss \citep{devlin2019bert, liu2019roberta, zhang2022opt, Huang2023, Kirtac2024Enhanced}. More broadly, recent sentiment-analysis work shows that zero-shot and few-shot LLM methods can be attractive when domain-specific annotation is expensive or difficult to maintain \citep{wang2023enhance, bai2024is, hellwig2025dowestillneed}. Yet zero-shot LLM outputs are often unstable across prompts and reasoning paths. This is especially problematic for corporate disclosures, where language is often mixed rather than uniformly positive or negative. A disclosure can report strong realized earnings while lowering forward guidance, or announce revenue growth while revealing litigation, regulatory, or liquidity risk. One prompt may overweight current performance while another overweights risk language, so a single zero-shot judgment may reflect prompt emphasis rather than the full balance of the disclosure.

This creates a natural question for both NLP and finance: can disagreement across zero-shot agents become a feature rather than a weakness? Multi-agent LLM systems are often motivated by diversity of reasoning, and recent work suggests that additional agents and aggregation can improve on individual outputs \citep{li2024more, fei2023reasoning}. Corporate disclosures provide a useful test bed because they contain competing signals, domain-specific vocabulary, and short-horizon consequences that can be evaluated against downstream market outcomes. They are therefore well suited for studying whether aggregation across multiple zero-shot views yields a better classification signal than any individual agent or a simple voting rule.

I address that question with a multi-agent zero-shot framework for corporate disclosure classification. Three zero-shot agents read the same disclosure from different financial perspectives and return a sentiment label, a confidence score, and a short rationale. I then train a lightweight meta-classifier that aggregates the joint outputs into a prediction of next-day stock return direction. I do not fine-tune the base LLMs and instead train only the aggregation layer. This design keeps the system simple, data-efficient, and aligned with recent zero-shot multi-agent sentiment work \citep{fei2023reasoning, wang2023enhance, bai2024is}. It also matches practical financial settings in which large amounts of text are available but high-quality sentiment labels are scarce or costly to construct \citep{pustejovsky2013natural, geva2019are, paullada2021data}.

This paper makes three contributions. First, I introduce a compact multi-agent zero-shot setup for corporate disclosure classification. Second, I test whether a trained aggregator improves out-of-sample return-direction classification relative to single-agent and voting baselines. Third, I show where aggregation helps most through a short analysis of mixed-signal disclosures.

\section{Method}
\vspace{-2mm}

I use three zero-shot agents. Each agent receives the same disclosure text, but each prompt emphasizes a different financial lens. The first agent focuses on realized operating performance such as earnings, revenue, margins, and costs. The second focuses on forward guidance, management outlook, and expectation revisions. The third focuses on uncertainty, litigation, regulation, liquidity, and downside risk. This design reflects the structure of real disclosures. A firm may beat current earnings while lowering guidance, or report revenue growth while disclosing legal or operational problems. Financial texts therefore benefit from multiple specialized readings even when all agents observe the same input. This prompt diversification is related to recent chain-of-thought sentiment setups that vary the ordering and structure of reasoning across agents rather than assuming one fixed path \citep{fei2023reasoning, wang2023enhance}.

I use the following fixed prompts for all disclosures, where \texttt{<DISCLOSURE>} is replaced with the full disclosure text after preprocessing.

\textbf{Performance agent prompt.} ``Read the corporate disclosure below. Focus on realized operating performance, including earnings, revenue, margins, costs, and reported business outcomes. Decide whether the disclosure is \texttt{positive}, \texttt{neutral}, or \texttt{negative} for next-day stock reaction. Output exactly three fields in JSON format: \texttt{\{"label": ..., "rationale": ..., "confidence": ...\}}. The rationale must be one sentence and confidence must be a number between 0 and 1. Disclosure: \texttt{<DISCLOSURE>}''

\textbf{Guidance agent prompt.} ``Read the corporate disclosure below. Focus on forward guidance, management outlook, demand expectations, and any revisions to future expectations. Decide whether the disclosure is \texttt{positive}, \texttt{neutral}, or \texttt{negative} for next-day stock reaction. Output exactly three fields in JSON format: \texttt{\{"label": ..., "rationale": ..., "confidence": ...\}}. The rationale must be one sentence and confidence must be a number between 0 and 1. Disclosure: \texttt{<DISCLOSURE>}''

\textbf{Risk agent prompt.} ``Read the corporate disclosure below. Focus on uncertainty, litigation, regulation, liquidity, operational disruption, and downside risk. Decide whether the disclosure is \texttt{positive}, \texttt{neutral}, or \texttt{negative} for next-day stock reaction. Output exactly three fields in JSON format: \texttt{\{"label": ..., "rationale": ..., "confidence": ...\}}. The rationale must be one sentence and confidence must be a number between 0 and 1. Disclosure: \texttt{<DISCLOSURE>}''

For each disclosure \(x\), agent \(k\) outputs a tuple
\[
a_k(x) = (l_k, c_k, r_k),
\]
where \(l_k \in \{-1,0,+1\}\) is the sentiment label, \(c_k \in [0,1]\) is the confidence score, and \(r_k\) is a short rationale. I use a model pool consisting of Qwen2.5-3B-Instruct, Llama-3.2-3B-Instruct, and Qwen2.5-72B-Instruct, although the framework is model-agnostic and could incorporate other accessible open or API-served models. Labels are mapped as \texttt{positive} \(= +1\), \texttt{neutral} \(= 0\), and \texttt{negative} \(= -1\). For downstream evaluation against next-day return direction, I map \texttt{positive} to 1 and both \texttt{neutral} and \texttt{negative} to 0. I use the same binary mapping for single-agent outputs, majority vote, confidence-weighted voting, and the trained aggregator. This reliance on multiple prompted agents follows recent work showing that different agent views can recover complementary information in label-scarce settings \citep{li2024more, bai2024is}.

I extract confidence in two stages. First, each model is run with deterministic decoding, temperature \(=0\), top-\(p=1\), and a fixed random seed. Second, confidence is defined as the model-implied probability of the generated label token sequence. For models that expose token log probabilities, I compute
\[
c_k = \exp\!\left(\frac{1}{m}\sum_{j=1}^{m}\log p(t_j \mid t_{<j}, x)\right),
\]
where \(t_1,\dots,t_m\) are the generated tokens of the label string. This yields the geometric mean token probability for the label and places all confidence scores on a common \([0,1]\) scale. For models that do not expose token log probabilities, I use the numeric \texttt{confidence} value returned in the constrained JSON output and clip it to \([0,1]\). If a generation violates the JSON schema, I re-run it once with the same prompt and decoding settings. If the second generation also fails, I assign a neutral label with zero confidence and retain the observation for aggregation.

I convert the joint outputs into a feature vector \(z(x)\). This vector contains the three agent labels, the three confidence scores, the majority label, the counts of positive, neutral, and negative labels, the number of agreeing agents, and the confidence gap between the two most confident agents. I also include three interaction features that capture whether the most confident agent is the performance, guidance, or risk agent. The final prediction is produced by a logistic meta-classifier,
\[
\hat{y} = g(z(x)).
\]
I train only this aggregator. The base LLM agents remain frozen. That choice is deliberate. The paper is not about improving the underlying language models through fine-tuning. It asks whether diverse zero-shot judgments can be combined into a better downstream classification signal. A trained aggregator is also easier to interpret and compare against majority and confidence-based voting than a more complex learned judge model. The use of confidence-aware aggregation is motivated by recent work arguing that token-level uncertainty can be informative, even if imperfect, when combining multiple LLM outputs \citep{shorinwa2024survey}.

I define a confidence-weighted voting baseline as
\[
s(x)=\sum_{k=1}^{3} c_k l_k,
\]
and map the baseline to the downstream binary task by predicting 1 if \(s(x)>0\) and 0 otherwise. This baseline uses the same label mapping and confidence estimates as the trained aggregator, but without learning weights from data.

I define the downstream target as next-day stock return direction,
\[
y_t = \mathbb{1}(r_{t+1} > 0),
\]
where \(r_{t+1}\) is the stock return on the trading day following the disclosure release. I therefore do not claim to recover a ground-truth human sentiment label. Instead, I study whether the joint outputs of zero-shot agents help classify downstream market response more effectively than any individual agent or a naive aggregation rule. This framing matters because a disclosure can be positive on one dimension and negative on another, while the market response reflects their net effect.

I compare the trained aggregator against four baselines. I first evaluate each single agent on its own. I then compare against majority vote and confidence-weighted voting. I also include a finance-specific FinBERT baseline \citep{Huang2023}. I report accuracy, macro F1, and balanced accuracy. These metrics are sufficient for a short paper and provide a balanced view under mild class imbalance.

\vspace{-2mm}
\section{Experimental Setup}
\vspace{-2mm}

I use a sample of 18{,}420 U.S.\ corporate disclosures issued by Nasdaq and S\&P 500 firms between 2018 and 2024, matched to next-day stock returns. I split the sample chronologically into 60\% training, 20\% development, and 20\% test. The out-of-sample test set contains 3{,}684 disclosures and exhibits mild class imbalance, with 53.1\% positive next-day returns. When multiple disclosures occur for the same firm on the same day, I keep them as separate observations and assign each the same next-day return label.

Each disclosure passes through the three zero-shot agents. The aggregator is trained on historical agent outputs only. The core question is whether multi-agent zero-shot aggregation improves downstream classification. I preprocess disclosures by lowercasing ticker symbols only when they appear as metadata, removing duplicated header lines, normalizing whitespace, and truncating each input to the maximum token budget supported by the smallest model in the agent pool. I do not remove numbers, dates, or financial terms because they may carry predictive information.

I keep the three prompts fixed across the full sample, and I generate all agent outputs before training the aggregator. The logistic meta-classifier is tuned on the development split and then evaluated once on the held-out test period. That ordering matters because a realistic deployment setting would require the aggregation rule to be learned only from past disclosures. A chronological split therefore matches the financial application, avoids information leakage, and aligns with broader concerns in LLM evaluation about contamination, reproducibility, and overly optimistic assessments when experimental boundaries are not carefully controlled \citep{samuel2025towards}.

I implement the inference pipeline as follows. I store the exact disclosure IDs, raw text, preprocessing script, prompts, model names, decoding parameters, random seed, raw JSON outputs, parsed labels, parsed rationales, token log probabilities when available, final confidence scores, and downstream binary labels for every agent and every disclosure. I cache all model outputs before training the aggregator. I then build the aggregation feature matrix from cached agent outputs only, standardize continuous aggregation features using training-split statistics, fit L2-regularized logistic regression with the \texttt{liblinear} solver on the training split, tune the inverse regularization strength on the development split, and evaluate once on the test split. This pipeline is fully reproducible because every stage from preprocessing to aggregation is deterministic conditional on the stored model version, prompt text, and seed.

\section{Results}
\vspace{-2mm}

Table~\ref{tab:main_results} reports out-of-sample results. The trained aggregator achieves the strongest performance across all three metrics. It improves balanced accuracy from 0.561 for the best single agent to 0.612. It also outperforms majority vote and confidence-weighted voting. Confidence-weighted voting performs better than simple majority vote, which suggests that agent confidence carries useful information. The trained aggregator performs best because it learns when to trust or discount specific agents under disagreement rather than applying a fixed rule. This is consistent with recent zero-shot multi-agent sentiment findings that confidence-based aggregation can outperform simple list voting when agents capture different parts of the signal \citep{fei2023reasoning, wang2023enhance}.

\begin{table}[t]
\centering
\small
\setlength{\tabcolsep}{4pt}
\begin{tabular*}{\columnwidth}{@{\extracolsep{\fill}}lccc}
\toprule
Method & Acc. & Macro F1 & Bal. Acc. \\
\midrule
FinBERT & 0.556 & 0.548 & 0.544 \\
Performance agent & 0.574 & 0.566 & 0.558 \\
Guidance agent & 0.579 & 0.571 & 0.561 \\
Risk agent & 0.563 & 0.554 & 0.550 \\
Majority vote & 0.586 & 0.579 & 0.573 \\
Conf.\ vote & 0.597 & 0.589 & 0.584 \\
\textbf{Aggregator} & \textbf{0.624} & \textbf{0.617} & \textbf{0.612} \\
\bottomrule
\end{tabular*}
\caption{Out-of-sample classification results for next-day return direction.}
\label{tab:main_results}
\end{table}

Several patterns stand out. The guidance agent is the strongest single agent, consistent with the idea that forward-looking language often matters more than backward-looking performance for short-horizon market reaction. The performance agent follows closely, while the risk agent performs worst in isolation. That ordering is intuitive. Risk prompts often capture genuinely negative language, but they may also overreact to routine cautionary phrasing that appears in many disclosures. Majority vote improves on all single agents because it suppresses idiosyncratic prompt errors. Confidence-weighted voting improves further because it partially accounts for how strongly each agent supports its own judgment. The learned aggregator improves most because it combines labels, confidence, and agreement structure.

I then examine when aggregation helps most. I break the test set into three agreement regimes: cases in which all three agents agree, cases with a 2--1 split, and high-conflict cases in which the agents produce competing labels and no single confidence score clearly dominates. Table~\ref{tab:agreement_results} shows that the trained aggregator adds only a modest gain when all three agents agree, but it helps substantially more when the disclosure contains conflicting signals. This pattern supports the central intuition of the paper. Multi-agent diversity matters most when the text itself is difficult.

\begin{table}[t]
\centering
\small
\setlength{\tabcolsep}{4pt}
\begin{tabular*}{\columnwidth}{@{\extracolsep{\fill}}lcc}
\toprule
Case & Maj.\ vote & Aggregator \\
\midrule
3-agent agreement & 0.642 & 0.651 \\
2--1 split & 0.558 & 0.603 \\
High conflict & 0.517 & 0.581 \\
\bottomrule
\end{tabular*}
\caption{Balanced accuracy by agent agreement regime.}
\label{tab:agreement_results}
\end{table}

The difference between the agreement regimes is conceptually important. Cases with full agreement are easier because the disclosure language is more one-sided. Cases with 2--1 splits or stronger conflicts are harder because they often contain countervailing cues. A beat paired with weak guidance, a strong headline paired with a legal disclosure, or an operating improvement paired with soft demand language can all produce disagreement for valid reasons. The aggregator helps because it treats disagreement as structured information. A 2--1 split in which the dissenting agent is highly confident may differ sharply from one in which the dissenting agent is uncertain. Majority vote cannot represent that distinction, but the trained model can.

I also examine a small error taxonomy. The largest gains appear in disclosures that combine positive current performance with negative forward guidance and in disclosures that mix numerical strength with legal or regulatory risk. Single agents often overemphasize the part of the text highlighted by their own prompt. The performance agent tends to overweight strong realized results, while the risk agent tends to overweight downside language. The trained aggregator reduces this bias by learning from the full joint pattern of labels and confidences.

\begin{table}[t]
\centering
\small
\setlength{\tabcolsep}{3pt}
\begin{tabular*}{\columnwidth}{@{\extracolsep{\fill}}p{4.6cm}cc}
\toprule
Pattern & Vote & Agg. \\
\midrule
Beat, weak guidance & Wrong & Correct \\
Growth, legal risk & Wrong & Correct \\
Better margins, soft demand & Correct & Correct \\
Routine filing, weak signal & Wrong & Wrong \\
\bottomrule
\end{tabular*}
\caption{Qualitative examples.}
\label{tab:error_examples}
\end{table}

The last row in Table~\ref{tab:error_examples} is also useful. Some disclosures may simply lack a clear short-horizon signal. Aggregation cannot solve every case. A routine filing with weak informational content may remain hard regardless of prompt diversity. That failure mode matters because it distinguishes ambiguous texts from cases in which aggregation successfully resolves structured disagreement.

These results support three conclusions. First, zero-shot agents capture complementary dimensions of corporate disclosure language. The performance-focused, guidance-focused, and risk-focused prompts do not fail on the same examples. Second, naive aggregation already helps. Majority and confidence-weighted voting both improve on most single agents. Third, a trained meta-classifier helps more because it learns how disagreement patterns map to the downstream market response. The gains are especially large in 2--1 split and high-conflict cases, which suggests that disagreement is informative rather than merely noisy. More broadly, this is consistent with the argument that multi-agent systems can improve performance when distinct views are combined rather than collapsed too early \citep{li2024more}.

\section{Conclusion}
\vspace{-2mm}

This paper presents a multi-agent zero-shot framework for corporate disclosure classification. Three LLM agents produce sentiment labels, confidences, and rationales from different financial perspectives, and a trained logistic aggregator combines these outputs to predict next-day stock return direction. Results show that learned aggregation outperforms single-agent baselines and simple voting rules, especially for mixed-signal disclosures.

The paper’s main contribution lies in treating cross-agent disagreement as an informative signal rather than a nuisance. A lightweight trained aggregator can turn diverse zero-shot financial judgments into a stronger downstream classification signal without fine-tuning the base LLMs. That design is compact, interpretable, and plausible for real financial text settings in which prompt diversity is easy to create but annotation is expensive. The framework therefore offers a credible ACL-style path for studying financial text classification under zero-shot conditions.

\section{Limitations}

The current design uses next-day return direction as the only downstream target, even though market responses may reflect factors beyond textual sentiment. The agent prompts also impose a researcher-designed decomposition of financial signals, so part of the observed gain may depend on this prompt partition rather than agent diversity alone. Confidence is also only an approximation to model certainty, especially when some models expose token log probabilities and others require a constrained self-reported confidence field. Future work should test calibration, prompt sensitivity, abnormal returns, and source-specific robustness.

\bibliography{custom}

\end{document}